\documentclass[a4paper,11pt]{article}
\usepackage{pos}
\usepackage{comment}

\def\nbslash{\rlap{\hspace{0.02cm}/}{\bar n}}
\def\nslash{\rlap{\hspace{0.02cm}/}{n}}
\newcommand{\eps}{\varepsilon}

\title{NLO analysis of the subleading-power $Q_1-Q_{7\gamma}$ interference in $\bar{B}\to X_s\gamma$ at large photon energies}
\ShortTitle{NLO analysis of the $Q_1-Q_{7\gamma}$ interference in $\bar{B}\to X_s\gamma$ at large photon energies}

\author*[a]{Riccardo Bartocci}
\author[b]{Philipp Böer}
\author[a]{Tobias Hurth}

\affiliation[a]{PRISMA+ Cluster of Excellence \& Mainz Institute for Theoretical Physics, \\ Johannes Gutenberg University, D-55099 Mainz, Germany}
\affiliation[b]{CERN, Theoretical Physics Department, CH-1211 Geneva 23, Switzerland}

\emailAdd{rbartocc@uni-mainz.de}
\emailAdd{philipp.boeer@cern.ch}
\emailAdd{hurth@uni-mainz.de}

\abstract{We report on progress towards including next-to-leading order (NLO) radiative corrections to the subleading-power factorization formula for the $Q_1^{q}-Q_{7\gamma}$ interference contribution in $\bar{B}\to X_s\gamma$ at large photon energies, $m_b - 2E_\gamma = \mathcal{O}(\Lambda_{\rm QCD})$. The novel ingredients for a NLO analysis are the one-loop anomalous dimension of the subleading shape function $g_{17}(\omega,\omega_1;\mu)$, which was recently computed by us, and the two-loop corrections to the jet function with an internal charm loop. Here we summarize the renormalization of $g_{17}(\omega,\omega_1;\mu)$, and further discuss that our insights allow for a simplified renormalization of a specific generalization of a three-particle $B$-meson light-cone distribution amplitude, but with non-aligned light-like separations of the fields.}

\FullConference{
European Physical Society Conference on High Energy Physics (EPS-HEP 2025)\\
7th July - 11th July, 2025\\
Marseille, France\\
\vspace{0.3cm}
MITP-25-063, CERN-TH-2025-197
}

\begin{document}
\maketitle
\section{Introduction}
Flavor-changing quark transitions provide highly sensitive probes of the Standard Model (SM). Their strong suppression within the SM makes them particularly suitable for revealing potential signals of new physics (NP). In this context, the inclusive penguin modes $\bar B \to X_{s,d} \gamma$ and $\bar B \to X_{s,d} \ell^+\ell^-$ are theoretically relatively clean channels for indirect NP searches~\cite{Hurth:2010tk}. However, to fully harness their discovery potential necessitates precise SM predictions, which demand in particular an improved understanding of the strong interactions dynamics in these decays. 

While perturbative calculations for $\bar{B} \to  X_s \gamma$ include corrections up to next-to-next-to leading order (NNLO), see e.g.~\cite{Misiak:2015xwa} and references therein, as well as multi-parton contributions at NLO~\cite{Brune:2025zhd}, the photon-energy spectrum is also sensitive to non-perturbative bound-state effects of the $\bar{B}$ meson. A separation of perturbative from non-perturbative dynamics can be achieved in an expansion in powers of $\Lambda_{\rm QCD}/m_b$ using Effective Field Theories (EFTs).

In this proceedings article, we study the inclusive decay $\bar{B} \rightarrow X_s \gamma$ near the kinematical endpoint, \mbox{$m_b - 2E_\gamma = \mathcal{O}(\Lambda_{\rm QCD}) \ll m_b$}. In this region, the hadronic state $X_s$ has large energy of $\mathcal{O}(m_b)$ but small invariant mass $\sim \mathcal{O}(m_b \Lambda_{\rm QCD}) \ll m_b^2$, which is still a perturbative scale. A local operator-product expansion fails to describe this jet-like configuration, but it can be analyzed with factorization methods from soft-collinear EFT (SCET)~\cite{Bauer:2000yr,Bauer:2001yt,Bauer:2002nz,Beneke:2002ph,Beneke:2002ni}. 
Systematic SCET analyses at LO in $\alpha_s$ within the endpoint region were presented 
in~\cite{Benzke:2010js,Hurth:2017xzf,Benzke:2017woq}.

Here we focus on the so-called single resolved photon contribution due to the interference between the weak operators $Q_1^q$ ($q = u,c$) and $Q_{7\gamma}$,
\begin{equation}
Q_1^q = [\bar{q} \gamma^\mu (1 - \gamma_5) b] [\bar{s} \gamma_\mu (1 - \gamma_5) q], \qquad
Q_{7\gamma} = \frac{-e m_b}{8\pi^2} \bar{s} \sigma_{\mu\nu} (1 + \gamma_5) F^{\mu\nu} b \,,
\end{equation}  
because it is the largest contribution among these nonlocal, $1/m_b$ power-suppressed terms. A factorization formula for this resolved contribution was presented at LO in~\cite{Benzke:2010js}, and takes the form 
\begin{equation}
\label{eq:facstructure}
 d\Gamma(\bar{B} \to X_s \gamma) \sim \int_0^1 \! du \, H(u;\mu) \int_{-\infty}^{\infty} \frac{d\omega_1}{\omega_1 + i0} \, \bar{J}(u,\omega_1;\mu) \int_{-\infty}^{\bar{\Lambda}} \! d\omega \, J(\omega;\mu) \, g_{17}(\omega,\omega_1;\mu) \,,
\end{equation}
with a perturbative hard matching coefficient $H(u) = 1 + \mathcal{O}(\alpha_s)$ that encapsulates physics at scales $\mu_h \sim m_b$, two perturbative jet functions $\bar{J}$ and $J$ with intrinsic scale $\mu_j \sim \sqrt{m_b \Lambda_{\rm QCD}}$, and the subleading shape function $g_{17}$ which parametrizes non-perturbative fluctuations at scales $\mu_s \sim \Lambda_{\rm QCD}$. For practical reasons we have taken out an explicit factor $1/\omega_1$ which belongs to the jet function $\bar{J}$. In~\cite{Hurth:2023paz} a failure of factorization in the resolved contribution $Q_8-Q_8$ was healed by using refactorization techniques~\cite{Boer:2018mgl,Liu:2020wbn,Beneke:2022obx,Bell:2022ott}.

Uncertainties from resolved-photon contributions are among the dominant ones in inclusive penguin modes. The $Q_1^c - Q_{7\gamma}$ interference gives $(5.45 \pm 2.55)\%$ 
\cite{Benzke:2020htm,Benzke:2023fmw} corresponding
to a range of $[2.9\%, 8\%]$. Here, the Voloshin term of $+3.3\%$ (using consistently LO Wilson coefficients), traditionally subtracted from the resolved contribution and computed via a local expansion~\cite{Grant:1997ec, Buchalla:1997ky, Ligeti:1997tc}, has been added back. In addition, a substantial scale ambiguity must be considered. At LO (including the Voloshin term), the scale of the hard functions, i.e. of the Wilson coefficients, is not fixed. Shifting their scale from the hard scale to the hard-collinear scale results in an increase of the range of more than $40\%$ to $[4.2\%,11.7\%]$.  The overall range including the scale ambiguity is then $[2.9\%,11.7\%]$. Here the charm-mass dependence of the Voloshin term is not taken into account yet.

Finally, we note that the effects of the resolved contributions still have to be described in terms of non-local operator matrix elements when the photon cut is moved outside the endpoint region~\cite{Benzke:2010js,Hurth:2017xzf,Benzke:2017woq}. In addition, the support properties of the shape functions imply that the resolved contributions -- besides the $Q_8-Q_8$ one -- are almost cut independent. Notice that experimental analyses typically impose a lower cut on the photon energies, $E_\gamma \gtrsim 1.7$ GeV, which is below the endpoint region considered here.

The significant uncertainties in the $Q_1^c - Q_{7\gamma}$ interference motivate a systematic calculation of $\mathcal{O}(\alpha_s)$ corrections within renormalization-group (RG) improved perturbation theory. Whereas the standard quark-jet function $J$ is known up to N$^3$LO~\cite{Bruser:2018rad} and one-loop corrections to the hard matching function $H$ can be extracted e.g. from~\cite{Beneke:2020vnb}, the missing required ingredients for a NLO analysis are $\mathcal{O}(\alpha_s)$ (two-loop) corrections to the jet function $\bar{J}$ with an internal charm loop, as well as the one-loop anomalous dimension of the shape function $g_{17}$. The latter was recently computed by us, and the corresponding RG equation was solved analytically~\cite{Bartocci:2024bbf}. In section~\ref{sec:g17} we summarize the main results, and section~\ref{sec:phiG} briefly discusses new insights from this analysis on the renormalization of a specific amplitude-level $B$-meson soft function with fields smeared on two different light cones.

\section{Renormalization of the subleading shape function}
\label{sec:g17}
The subleading shape function $g_{17}$ is defined via a forward matrix element between two static $\bar{B}$-meson states~\cite{Benzke:2010js,Bartocci:2024bbf}, 
\begin{align}\label{eq:g17def}
   g_{17}(\omega,\omega_1;\mu) = \frac{1}{2M_B} \int\frac{dr}{2\pi}\,e^{-i\omega_1 r}
    \int\frac{dt}{2\pi}\,e^{-i\omega t} \, \langle\bar B_v| \mathcal{O}_{17}(t,r) |\bar B_v\rangle \,,
\end{align}
of an operator in Heavy-Quark Effective Theory (HQET),
\begin{equation}
\label{eq:Q17}
    \mathcal{O}_{17}(t,r) = \big(\bar h_v S_n\big)_-(tn)\,
    \nbslash \big(S_n^\dagger S_{\bar n}\big)_+(0)\,
    i\gamma_\alpha^\perp\bar n_\beta\,
    \big(S_{\bar n}^\dagger\,g_s G_s^{\alpha\beta} S_{\bar n} 
    \big)_+(r\bar n)\,
    \big(S_{\bar n}^\dagger h_v\big)_+(0) \,.
\end{equation}
Here $\bar{n}^\mu = p_\gamma^\mu/E_\gamma$ is a light-like vector in the direction of the photon momentum, and $n^\mu$ points in the opposite direction of the collimated and energetic hadronic $X_s$ system, such that $n^2 = \bar{n}^2 = 0$ and $n \cdot \bar{n} = 2$.
The $S_n$ and $S_{\bar{n}}$ are soft Wilson lines that combine to finite segments and ensure gauge invariance of the non-local operator. Lastly, the subscripts ``$+$'' (``$-$'') label fields that belong to the amplitude (complex conjugate amplitude), and contractions of fields with different indices are defined with on-shell cut propagators. The evaluation of such matrix elements can also be formulated at the level of the path integral using the Keldysh formalism~\cite{Schwinger:1960qe,Keldysh:1964ud}.

A complication of the operator $\mathcal{O}_{17}(t,r)$, compared to more standard light-like correlators, arises from the fact that its fields are smeared along two distinct light-cones $n^\mu$ and $\bar{n}^\mu$. The renormalization of such a ``multi-light-cone'' soft function relevant to $B$ decays was first studied in~\cite{Beneke:2022msp} for the $B$-meson light-cone distribution amplitude (LCDA) including QED effects. Generalizations of LCDAs with fields on different light-cones recently attracted some attention in the context of power-corrections and QED corrections to various exclusive $b$-hadron decays~\cite{Beneke:2019slt,Beneke:2020vnb,Beneke:2022msp,Qin:2022rlk,Piscopo:2023opf,Huang:2023jdu,Boer:2023vsg,Feldmann:2023plv}.

\begin{figure}[t]
    \centering
    \includegraphics[width=1.0\textwidth]{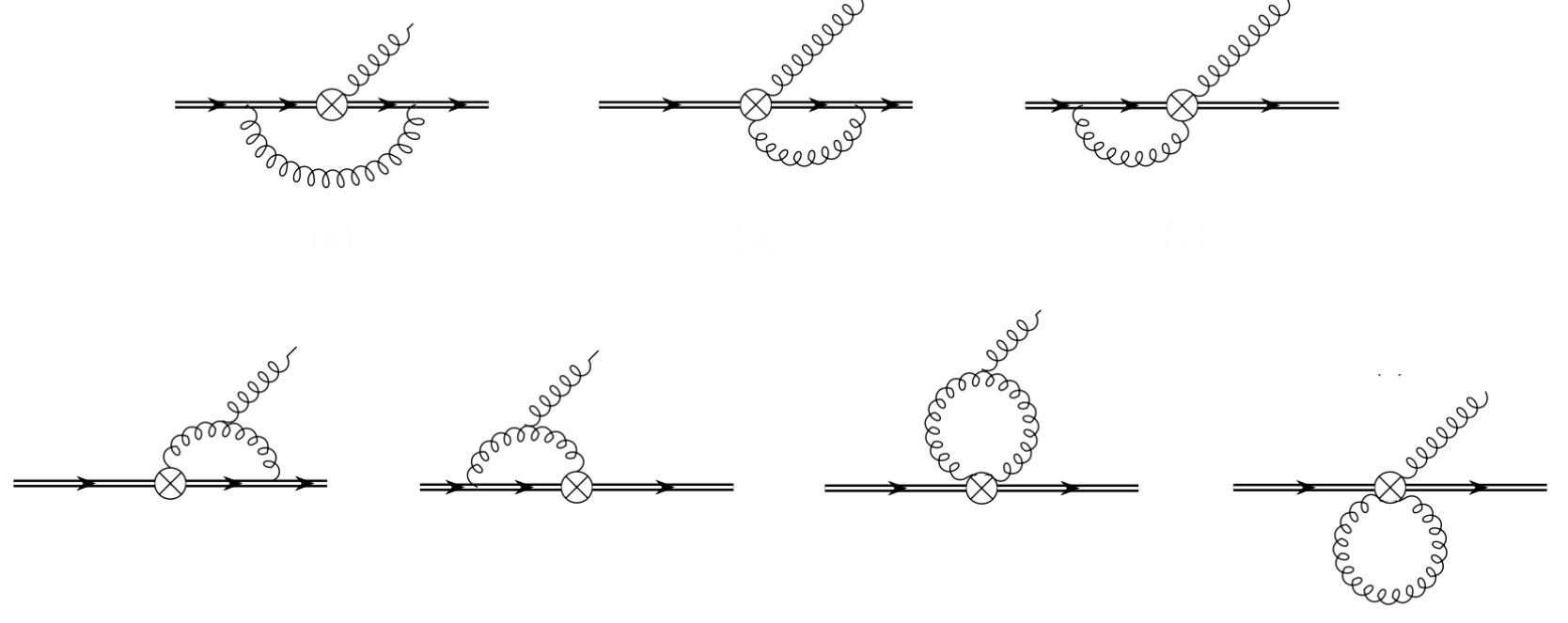}
    \caption{One-loop diagrams contributing to the UV singularities of the HQET operator $\mathcal{O}_{17}(t,r)$. The effective vertex denoted by the $\otimes$ symbol represents the non-local structure of fields and Wilson lines in~\eqref{eq:Q17}.}
    \label{fig:g17_diagrams}
\end{figure}

We compute the anomalous dimension of $\mathcal{O}_{17}(t,r)$ by choosing partonic external states, i.e. an incoming heavy-quark state, and an outgoing heavy quark and gluon state. The one-loop diagrams contributing to the ultra-violet (UV) singularities are shown in Fig.~\ref{fig:g17_diagrams}, and Feynman rules for (multiple) gluon emissions from the effective vertex are presented in~\cite{Bartocci:2024bbf}. Loop integrals are evaluated in $d = 4 - 2\eps$ space-time dimensions while keeping external legs off-shell to regulate infra-red (IR) divergences. The expansion around $\eps = 0$ has to be performed in the distribution sense, and besides standard plus-distributions,
\begin{equation}
    \int \! d\omega' \, \big [ \dots \big]_+ \, f(\omega') \equiv \int \! d\omega' \, \big[ \dots \big] \, (f(\omega')-f(\omega)) \,,
\end{equation}
also the modified plus-distributions,
\begin{equation}
    \int \! d\omega' \, \big[ \dots \big]_{\oplus/\ominus} \, f(\omega') \equiv \int \! d\omega' \, \big[ \dots \big] \, (f(\omega')-\theta(\pm \omega')f(\omega)) \,,
\end{equation}
arise as a consequence of positive and negative light-cone momenta $\omega$ and $\omega_1$ (cf. discussion in~\cite{Beneke:2022msp}). Furthermore, defining 
\begin{eqnarray}
\label{eq:FandGdistributions}
F^{>}(\omega, \omega^{\prime}) & = &
\left [ \frac{\omega \, \theta (\omega^{\prime} - \omega)} {\omega^{\prime} \,  (\omega^{\prime} - \omega)}   \right ]_{+}
+ \left [ \frac{\theta (\omega - \omega^{\prime})}  {\omega - \omega^{\prime}}  \right ]_{\oplus} \,,
\nonumber \\
F^{<}(\omega, \omega^{\prime})  & = &
\left [ \frac{\omega \, \theta (\omega - \omega^{\prime})} {\omega^{\prime} \,  (\omega- \omega^{\prime})}   \right ]_{+}
+ \left [ \frac{\theta (\omega^{\prime} - \omega)}  {\omega^{\prime} - \omega}  \right ]_{\ominus}  \,,
\nonumber \\
G^{>}(\omega, \omega^{\prime})  & = & (\omega + \omega^{\prime}) \,
\left [ \frac{\theta (\omega^{\prime} - \omega)} {\omega^{\prime} \,  (\omega^{\prime} - \omega)}   \right ]_{+}
- i  \pi  \delta(\omega - \omega^{\prime}) \,,
\nonumber \\
G^{<}(\omega, \omega^{\prime})  & = &   (\omega + \omega^{\prime}) \,
\left [ \frac{ \theta (\omega - \omega^{\prime})} {\omega^{\prime} \,  (\omega- \omega^{\prime})}   \right ]_{+}
+ i  \pi \delta(\omega - \omega^{\prime})  \,,
\end{eqnarray}
the following linear combinations of plus- and modified plus-distributions will arise~\cite{Beneke:2022msp}
\begin{equation}
\label{eq:H_distributions}
H_{\pm}(\omega, \omega^{\prime}) =  \theta( \pm \omega)  F^{> (<)} (\omega, \omega^{\prime})
+  \theta( \mp \omega)  G^{< (>)} (\omega, \omega^{\prime}) \,,
\end{equation}
which obey~\cite{Bartocci:2024bbf}
\begin{equation}
\label{eq:Hpmidentity}
  H_+(\omega,\omega') - H_-(\omega,\omega') = \frac{1}{\omega - \omega' - i0} \,.
\end{equation}

Evaluating the UV-divergent contributions from the diagrams in Fig.~\ref{fig:g17_diagrams}, we find for the one-loop anomalous dimension the structure
\begin{align}
\label{eq:gamma17}
 \gamma_{17}(\omega,\omega_1,\omega',\omega'_1;\mu)
 = \frac{\alpha_s}{\pi} \Big\{ C_F \, \delta(\omega_1 - \omega'_1) \gamma_{n}(\omega,\omega';\mu)
+ \frac{C_A}{2} \, \delta(\omega - \omega') \gamma_{\bar{n}}(\omega_1,\omega'_1;\mu) \Big\} \,.
\end{align}
Here the Abelian piece
\begin{equation}
\label{eq:gamman}
    \gamma_{n}(\omega,\omega';\mu) = \left(\ln \frac{\mu^2}{\omega^2} - 1\right) \delta(\omega-\omega')
    - 2\theta(\omega) \left[\frac{\theta(\omega'-\omega)}{\omega'(\omega'-\omega)}\right]_+ \omega' 
    - 2\theta(-\omega) \left[\frac{\theta(\omega'-\omega)}{\omega'-\omega}\right]_\ominus
\end{equation}
coincides with the well-known scale evolution of the leading shape function~\cite{Bosch:2004th},\footnote{Notice that~\eqref{eq:gamman} was derived \emph{without} assuming an upper cut-off $\bar{\Lambda} = m_B - m_b$ for the $\omega$ integral, cf.~\eqref{eq:facstructure}.} and the non-Abelian piece reads
\begin{equation}
\label{eq:gammanbar}
    \gamma_{\bar{n}}(\omega_1,\omega'_1;\mu) = \ln \frac{\mu^2}{\omega_1^2} \, \delta(\omega_1-\omega'_1) - \text{Re} \, \mathbf{H}(\omega_1,\omega'_1) 
    + \frac{2 \omega_1}{(\omega'_1)^2} \big[ \theta(\omega_1)\theta(\omega'_1 - \omega_1) - \theta(-\omega_1) \theta(\omega_1 - \omega'_1) \big] \,,
\end{equation}
with $\mathbf{H}(\omega_1,\omega'_1) = H_+(\omega_1,\omega'_1) + H_-(\omega_1,\omega'_1)$. Importantly, while $\gamma_{n}$ only acts on the soft momenta associated with the $n^\mu$ light-cone direction, $\gamma_{\bar{n}}$ only acts on momenta associated with the $\bar{n}^\mu$ direction. This decoupling of the two sectors is a manifestation of soft-collinear factorization, and to obtain the result~\eqref{eq:gamma17} it is crucial that the diagrams are evaluated with the respective cuts. However, a similar observation can also be made for soft functions at the amplitude level by using different arguments, as will be mentioned in section~\ref{sec:phiG}. 

One consequence of this decoupling is that the RG equation reduces to two individual equations that can be solved independently. The solution takes the factorized form
\begin{equation} \label{eq:g17_evolution}
    {g}_{17}(\omega,\omega_1;\mu) = \int_\omega^{\Bar{\Lambda}}\frac{d\omega'}{\omega'-\omega} \, U_n^{(17)}(\omega,\omega';\mu,\mu_0) \int_{-\infty}^{\infty} \frac{d\omega'_1}{\big\vert \omega'_1 \big\vert} \, U_{\bar{n}}^{(17)}(\omega_1,\omega'_1;\mu,\mu_0) \, g_{17}(\omega',\omega_1^\prime;\mu_0) \,,
\end{equation}
with the evolution function in the $n^\mu$ sector 
\begin{equation}
   U_n^{(17)}(\omega,\omega';\mu,\mu_0) = \frac{e^{2V + 2\gamma_E a}}{\Gamma(-2a)} \, \left(\frac{\mu_0}{\omega'-\omega}\right)^{\!2a}\,,
\end{equation}
with 
\begin{align}
    V(\mu,\mu_0) = -\int_{\mu_0}^\mu \frac{d\mu'}{\mu'} \frac{\alpha_s(\mu') C_F } {\pi} \bigg[  \ln \frac{\mu'}{\mu_0} -\frac12  \bigg] \;, \qquad  
    a(\mu,\mu_0) = -\int_{\mu_0}^\mu \frac{d\mu'}{\mu'} \frac{\alpha_s(\mu')C_F}{\pi} \,.
\end{align}
The evolution function in the $\bar{n}^\mu$ sector is more complicated, and can be expressed in terms of a hypergeometric function $_2F_1$ as well as a Meijer-G function. Defining $\tau = \omega'_1/\omega_1$, the result reads
\begin{align} 
\label{eq:CAevolutionfunction}
     U_{\bar{n}}^{(17)}(\omega_1,\omega'_1;\mu,\mu_0) 
     = \, &-e^{V_{1}+2\gamma_E a_{1}} \left(\frac{\mu_0}{\vert \omega'_1 \vert}\right)^{\! a_1} \bigg\{ 
     \theta(\tau) G^{1,2}_{3,3} \bigg( 
     \begin{array}{cccc} -1, \!\!  &1, \!\!  &a_1/2 \\  a_1+1, \!\!  &a_1-1, \!\! &a_1/2 \end{array}
     \bigg| \, \tau \bigg) \\[0.2em]
     + \, \frac{1}{2 \pi} \sin&\left(\frac{a_1 \pi}{2} \right)\theta(-\tau) 
     \Gamma(1+a_1)  \Gamma(3+a_1) (-\tau)^{1+a_1} {}_2F_1(1+a_1,3+a_1,3;\tau) \bigg\} \,, \nonumber 
\end{align}
with evolution factors
\begin{align}
    V_1(\mu,\mu_0) = -\int_{\mu_0}^\mu \frac{d\mu'}{\mu'} \frac{\alpha_s(\mu') C_A }{\pi}   \ln \frac{\mu'}{\mu_0} \;, \qquad  
    a_1(\mu,\mu_0) = -\int_{\mu_0}^\mu \frac{d\mu'}{\mu'} \frac{\alpha_s(\mu')C_A}{\pi} \,.
\end{align}
Given a certain phenomenological model for $g_{17}(\omega,\omega_1;\mu_0)$ at some low scale $\mu_0 \sim \mu_s \sim \Lambda_{\rm QCD}$, these expressions can be used to resum parametrically large logarithms between $\mu_s$ and the hard-collinear scale $\mu_j \sim \sqrt{m_b \Lambda_{\rm QCD}}$. Furthermore, studying the asymptotic behavior for large (positive and negative) momenta shows the existence of the relevant convolution integrals in~\eqref{eq:facstructure}.

\section{A three-particle \texorpdfstring{$B$}{B}-meson LCDA with fields smeared on two light-cones}
\label{sec:phiG}
An amplitude-level analogue to $g_{17}$ was introduced in~\cite{Qin:2022rlk} in the context of rare exclusive $\bar{B}_{s,d} \to \gamma \gamma$ decays. Its definition reads
\begin{align}
{\cal F}_B(\mu) \Phi_{\rm G}(\omega, \omega_1; \mu) = \frac{1}{2M_B} \int\frac{dr}{2\pi} \, e^{i\omega_1 r} \int\frac{dt}{2\pi}\,e^{i\omega t} \, \langle 0 | \mathcal{O}_G(t,r) | \bar B_v \rangle \,,
\label{eq:phi_G}
\end{align}
with the static $B$-meson decay constant ${\cal F}_B(\mu)$, and the HQET operator
\begin{equation}
\label{eq:exclusive_OG}
 \mathcal{O}_G(t,r) = (\bar q_{s} S_{n})(t n) \nslash (S_{n}^{\dagger} S_{\bar n})(0) \, \gamma^{\perp}_{\alpha} \bar n_{\beta} \gamma_5 \,
(S_{\bar n}^{\dagger} \, g_s G_s^{\alpha\beta} S_{\bar n} )(r \bar n) \,
(S_{\bar n}^{\dagger} h_v)(0) \,.
\end{equation}
A peculiar feature of exclusive $B$-meson soft functions with fields along two light-cone directions is that their support extends to the entire real axis, i.e. these functions do not vanish for negative arguments~\cite{Beneke:2022msp}. For the case of $\Phi_G(\omega,\omega_1;\mu)$, this applies to both variables: $\omega, \omega_1 \in (-\infty,\infty)$. The anomalous dimension again contains the distributions $H_+$ and $H_-$~\cite{Huang:2023jdu},
\begin{align}
\label{eq:exclusive_gamma} \Gamma_G(\omega,\omega_1,\omega',\omega'_1;\mu) = \, &\frac{\alpha_s C_F}{\pi}
\bigg[ \left( \ln{\frac{\mu}{\omega-i0}} -\frac{1}{2} \right) \delta(\omega - \omega^{\prime}) -  H_{+}(\omega, \omega^{\prime}) \bigg] \delta(\omega_1 - \omega_1^{\prime}) \nonumber \\[0.2em]
+ &\frac{\alpha_s C_A}{\pi} \bigg[ \left( \ln{\frac{\mu}{\omega_1-i0}} + \frac{i\pi}{2} \right) \delta(\omega_1 - \omega'_1) -  H_{+}(\omega_1, \omega'_1) \nonumber \\[0.2em]
&\phantom{\frac{\alpha_s C_A}{\pi}} \, +  \frac{\omega_1}{(\omega'_1)^2}
\left[ \theta(\omega_1) \theta(\omega'_1 - \omega_1) - \theta(-\omega_1) \theta(\omega_1 - \omega'_1) \right] \bigg] \delta(\omega - \omega') \nonumber \\[0.2em]
+ &\frac{\alpha_s}{\pi}\, \Delta\Gamma_G(\omega,\omega_1;\omega',\omega'_1;\mu) \,,
\end{align}
with
\begin{align}
\label{eq:exclusive_Deltagamma}
\Delta\Gamma_G = {\frac{i}{4}} {\frac{C_A}{\pi}}
\left[ \Delta H (\omega, \omega^{\prime}) 
- 2 i \pi \delta(\omega -  \omega^{\prime}) \right]
 \left[\Delta H(\omega_1, \omega_1^{\prime}) 
- 2 i \pi \delta(\omega_1 - \omega'_1)  \right] \,,
\end{align}
where $\Delta H = H_+ - H_-$. Notice that the distribution $H_-$ only appears in the term $\Delta \Gamma_G$.

In contrast to~\eqref{eq:gamma17}, the structure in~\eqref{eq:exclusive_Deltagamma} is non-local in both light-cone directions and superficially violates the decoupling of the two sectors, which might hint at additional factorization-breaking modes. However, this issue was resolved in~\cite{Bartocci:2024bbf} by the following observations: The function $\Phi_{\rm G}(\omega, \omega_1; \mu)$ is supported on the entire real axis, and is convoluted in a factorization theorem with jet functions that are analytic in the lower half of the complex $\omega$ and $\omega_1$ planes. This is a manifestation of causality and the large-energy flow through the respective Feynman diagrams. Due to the identity~\eqref{eq:Hpmidentity} and the location of the singularity of the linear propagator on its right-hand side, the two distributions $H_+$ and $H_-$ can be identified, and the contribution $\Delta \Gamma_G$ collapses to a simple local phase term on the physically relevant function space, 
\begin{equation}
 \frac{\alpha_s}{\pi} \Delta\Gamma_G \to -\frac{\alpha_s C_A}{\pi} i \pi \delta(\omega-\omega^\prime)\delta(\omega_1-\omega_1^\prime) \,.
\end{equation}
Exploiting the analytic properties leads to the so-called ``reduced anomalous dimension''~\cite{Beneke:2022msp,Bartocci:2024bbf},
\begin{equation}
\label{eq:exclusive_anomdim_factorisation}
\Gamma_{G}(\omega,\omega_1,\omega',\omega'_1;\mu)
 = \frac{\alpha_s}{\pi} \Big\{ C_F \, \delta(\omega_1 - \omega'_1) \Gamma_{n}(\omega,\omega';\mu)
+ C_A \, \delta(\omega - \omega') \Gamma_{\bar{n}}(\omega_1,\omega'_1;\mu) \Big\} \,,
\end{equation}
in which the two sectors decouple, and soft-collinear factorization is manifest. Furthermore, because the distribution $H_-$ does no longer appear in~\eqref{eq:exclusive_anomdim_factorisation}, a positively-supported function at the low scale $\mu_0$ does \emph{not} acquire negative support at $\mu > \mu_0$ in either of the two variables $\omega$ or $\omega_1$ under RG evolution with the reduced anomalous dimension.

We here refrain from repeating the individual terms $\Gamma_{n}$ and $\Gamma_{\bar{n}}$ and refer the reader to~\cite{Bartocci:2024bbf}, where also the analytic solutions to the RG equations and a discussion of the asymptotic behavior of $\Phi_{\rm G}(\omega, \omega_1; \mu)$ can be found.

\section{Conclusion and outlook}
\label{sec:conclusion}
Recent progress towards including next-to-leading order (NLO) radiative corrections to the single resolved photon $Q_1^q - Q_{7\gamma}$ interference contribution in $\bar{B} \to X_s \gamma$ is presented. Once completed, we expect our results to reduce the significant uncertainties from scale ambiguities and charm mass dependence in this contribution.

In this proceedings article, we primarily discuss the renormalization of the subleading shape function $g_{17}(\omega,\omega_1;\mu)$, which enters the factorization formula of this power-suppressed contribution as non-perturbative input. Our insights from this renormalization study of an amplitude-squared soft function beyond leading power also provide an improved understanding of amplitude-level soft functions with fields smeared on different light cones. The latter have recently attracted some attention in the literature, and might shed some new light on charm-loop contributions in $b \to s\gamma$ and $b \to s \ell \ell$ induced channels. 

The last missing piece to complete the NLO calculation are the two-loop corrections to the jet function with an internal charm loop. Results in the massless-quark limit, $m_c \to m_u = 0$, will be presented in an upcoming publication~\cite{toappear}, and they provide non-trivial cross checks for the consistency of the factorization framework. For phenomenological applications these results need to be generalized to $m_c \neq 0$, which is currently work in progress.

\section*{Acknowledgments} 
 R.B. thanks the organizers and conveners of EPS 2025 for an outstanding conference. This work is supported by  the  Cluster  of  Excellence  ``Precision  Physics,  Fundamental
Interactions, and Structure of Matter" (PRISMA$^+$ EXC 2118/1) funded by the German Research Foundation (DFG) within the German Excellence Strategy (Project ID 390831469). The research of P.B. was funded by the European Union's Horizon 2020 Research and Innovation Program under the Marie Sk\l{}odowska-Curie grant
agreement No.101146976.

\bibliographystyle{JHEP}
\bibliography{bibliography}

\providecommand{\href}[2]{#2}\begingroup\raggedright\begin{thebibliography}{10}

\bibitem{Hurth:2010tk}
T.~Hurth and M.~Nakao, \emph{{Radiative and Electroweak Penguin Decays of B Mesons}}, \href{https://doi.org/10.1146/annurev.nucl.012809.104424}{\emph{Ann. Rev. Nucl. Part. Sci.} {\bfseries 60} (2010) 645--677}, [\href{https://arxiv.org/abs/1005.1224}{{\ttfamily 1005.1224}}].

\bibitem{Misiak:2015xwa}
M.~Misiak et~al., \emph{{Updated NNLO QCD predictions for the weak radiative B-meson decays}}, \href{https://doi.org/10.1103/PhysRevLett.114.221801}{\emph{Phys. Rev. Lett.} {\bfseries 114} (2015) 221801}, [\href{https://arxiv.org/abs/1503.01789}{{\ttfamily 1503.01789}}].

\bibitem{Brune:2025zhd}
K.~Brune, T.~Huber and L.-T. Moos, \emph{{Multi-parton contributions to $\bar B \to X_s \gamma$ at NLO}},  \href{https://arxiv.org/abs/2509.22564}{{\ttfamily 2509.22564}}.

\bibitem{Bauer:2000yr}
C.~W. Bauer, S.~Fleming, D.~Pirjol and I.~W. Stewart, \emph{{An Effective field theory for collinear and soft gluons: Heavy to light decays}}, \href{https://doi.org/10.1103/PhysRevD.63.114020}{\emph{Phys. Rev. D} {\bfseries 63} (2001) 114020}, [\href{https://arxiv.org/abs/hep-ph/0011336}{{\ttfamily hep-ph/0011336}}].

\bibitem{Bauer:2001yt}
C.~W. Bauer, D.~Pirjol and I.~W. Stewart, \emph{{Soft collinear factorization in effective field theory}}, \href{https://doi.org/10.1103/PhysRevD.65.054022}{\emph{Phys. Rev. D} {\bfseries 65} (2002) 054022}, [\href{https://arxiv.org/abs/hep-ph/0109045}{{\ttfamily hep-ph/0109045}}].

\bibitem{Bauer:2002nz}
C.~W. Bauer, S.~Fleming, D.~Pirjol, I.~Z. Rothstein and I.~W. Stewart, \emph{{Hard scattering factorization from effective field theory}}, \href{https://doi.org/10.1103/PhysRevD.66.014017}{\emph{Phys. Rev. D} {\bfseries 66} (2002) 014017}, [\href{https://arxiv.org/abs/hep-ph/0202088}{{\ttfamily hep-ph/0202088}}].

\bibitem{Beneke:2002ph}
M.~Beneke, A.~P. Chapovsky, M.~Diehl and T.~Feldmann, \emph{{Soft collinear effective theory and heavy to light currents beyond leading power}}, \href{https://doi.org/10.1016/S0550-3213(02)00687-9}{\emph{Nucl. Phys. B} {\bfseries 643} (2002) 431--476}, [\href{https://arxiv.org/abs/hep-ph/0206152}{{\ttfamily hep-ph/0206152}}].

\bibitem{Beneke:2002ni}
M.~Beneke and T.~Feldmann, \emph{{Multipole expanded soft collinear effective theory with nonAbelian gauge symmetry}}, \href{https://doi.org/10.1016/S0370-2693(02)03204-5}{\emph{Phys. Lett. B} {\bfseries 553} (2003) 267--276}, [\href{https://arxiv.org/abs/hep-ph/0211358}{{\ttfamily hep-ph/0211358}}].

\bibitem{Benzke:2010js}
M.~Benzke, S.~J. Lee, M.~Neubert and G.~Paz, \emph{{Factorization at Subleading Power and Irreducible Uncertainties in $\bar B\to X_s\gamma$ Decay}}, \href{https://doi.org/10.1007/JHEP08(2010)099}{\emph{JHEP} {\bfseries 08} (2010) 099}, [\href{https://arxiv.org/abs/1003.5012}{{\ttfamily 1003.5012}}].

\bibitem{Hurth:2017xzf}
T.~Hurth, M.~Fickinger, S.~Turczyk and M.~Benzke, \emph{{Resolved Power Corrections to the Inclusive Decay $\bar B \to X_s \ell^+\ell^-$}}, \href{https://doi.org/10.1016/j.nuclphysbps.2017.03.011}{\emph{Nucl. Part. Phys. Proc.} {\bfseries 285-286} (2017) 57--62}, [\href{https://arxiv.org/abs/1711.01162}{{\ttfamily 1711.01162}}].

\bibitem{Benzke:2017woq}
M.~Benzke, T.~Hurth and S.~Turczyk, \emph{{Subleading power factorization in $ \bar{B}\to {X}_s{\ell}^{+}{\ell}^{-} $}}, \href{https://doi.org/10.1007/JHEP10(2017)031}{\emph{JHEP} {\bfseries 10} (2017) 031}, [\href{https://arxiv.org/abs/1705.10366}{{\ttfamily 1705.10366}}].

\bibitem{Hurth:2023paz}
T.~Hurth and R.~Szafron, \emph{{Refactorisation in subleading $\bar{B} \to X_s \gamma$}}, \href{https://doi.org/10.1016/j.nuclphysb.2023.116200}{\emph{Nucl. Phys. B} {\bfseries 991} (2023) 116200}, [\href{https://arxiv.org/abs/2301.01739}{{\ttfamily 2301.01739}}].

\bibitem{Boer:2018mgl}
P.~B\"oer, \emph{{QCD Factorisation in Exclusive Semileptonic B Decays New Applications and Resummation of Rapidity Logarithms}}, Ph.D. thesis, Siegen U., 2018.

\bibitem{Liu:2020wbn}
Z.~L. Liu, B.~Mecaj, M.~Neubert and X.~Wang, \emph{{Factorization at subleading power and endpoint divergences in $h\to\gamma\gamma$ decay. Part II. Renormalization and scale evolution}}, \href{https://doi.org/10.1007/JHEP01(2021)077}{\emph{JHEP} {\bfseries 01} (2021) 077}, [\href{https://arxiv.org/abs/2009.06779}{{\ttfamily 2009.06779}}].

\bibitem{Beneke:2022obx}
M.~Beneke, M.~Garny, S.~Jaskiewicz, J.~Strohm, R.~Szafron, L.~Vernazza et~al., \emph{{Next-to-leading power endpoint factorization and resummation for off-diagonal \textquotedblleft{}gluon\textquotedblright{} thrust}}, \href{https://doi.org/10.1007/JHEP07(2022)144}{\emph{JHEP} {\bfseries 07} (2022) 144}, [\href{https://arxiv.org/abs/2205.04479}{{\ttfamily 2205.04479}}].

\bibitem{Bell:2022ott}
G.~Bell, P.~B\"oer and T.~Feldmann, \emph{{Muon-electron backward scattering: a prime example for endpoint singularities in SCET}}, \href{https://doi.org/10.1007/JHEP09(2022)183}{\emph{JHEP} {\bfseries 09} (2022) 183}, [\href{https://arxiv.org/abs/2205.06021}{{\ttfamily 2205.06021}}].

\bibitem{Benzke:2020htm}
M.~Benzke and T.~Hurth, \emph{{Resolved $1/m_b$ contributions to $\bar B \to X_{s,d} \ell^+\ell^-$ and $\bar B \to X_s \gamma$}}, \href{https://doi.org/10.1103/PhysRevD.102.114024}{\emph{Phys. Rev. D} {\bfseries 102} (2020) 114024}, [\href{https://arxiv.org/abs/2006.00624}{{\ttfamily 2006.00624}}].

\bibitem{Benzke:2023fmw}
M.~Benzke and T.~Hurth, \emph{{Recent progress on the nonlocal power corrections to the inclusive penguin decays B {\textrightarrow} X s {\ensuremath{\gamma}} and B {\textrightarrow} X s {\ensuremath{\ell}}{\ensuremath{\ell}}}},  in \emph{{Strong interactions from QCD to new strong dynamics at LHC and Future Colliders}}, pp.~101--108, 3, 2023, \href{https://arxiv.org/abs/2303.06447}{{\ttfamily 2303.06447}}.

\bibitem{Grant:1997ec}
A.~K. Grant, A.~G. Morgan, S.~Nussinov and R.~D. Peccei, \emph{{Comment on nonperturbative $O(1/m_c^2)$ corrections to $\Gamma(\bar{B} \to X_s \gamma)$}}, \href{https://doi.org/10.1103/PhysRevD.56.3151}{\emph{Phys. Rev. D} {\bfseries 56} (1997) 3151--3154}, [\href{https://arxiv.org/abs/hep-ph/9702380}{{\ttfamily hep-ph/9702380}}].

\bibitem{Buchalla:1997ky}
G.~Buchalla, G.~Isidori and S.~J. Rey, \emph{{Corrections of order $\Lambda_{QCD}^2 / m_c^2$ to inclusive rare B decays}}, \href{https://doi.org/10.1016/S0550-3213(97)00674-3}{\emph{Nucl. Phys. B} {\bfseries 511} (1998) 594--610}, [\href{https://arxiv.org/abs/hep-ph/9705253}{{\ttfamily hep-ph/9705253}}].

\bibitem{Ligeti:1997tc}
Z.~Ligeti, L.~Randall and M.~B. Wise, \emph{{Comment on nonperturbative effects in $\bar{B} \to X_s\gamma$}}, \href{https://doi.org/10.1016/S0370-2693(97)00304-3}{\emph{Phys. Lett. B} {\bfseries 402} (1997) 178--182}, [\href{https://arxiv.org/abs/hep-ph/9702322}{{\ttfamily hep-ph/9702322}}].

\bibitem{Bruser:2018rad}
R.~Br{\"u}ser, Z.~L. Liu and M.~Stahlhofen, \emph{{Three-Loop Quark Jet Function}}, \href{https://doi.org/10.1103/PhysRevLett.121.072003}{\emph{Phys. Rev. Lett.} {\bfseries 121} (2018) 072003}, [\href{https://arxiv.org/abs/1804.09722}{{\ttfamily 1804.09722}}].

\bibitem{Beneke:2020vnb}
M.~Beneke, P.~B{\"o}er, J.-N. Toelstede and K.~K. Vos, \emph{{QED factorization of non-leptonic $B$ decays}}, \href{https://doi.org/10.1007/JHEP11(2020)081}{\emph{JHEP} {\bfseries 11} (2020) 081}, [\href{https://arxiv.org/abs/2008.10615}{{\ttfamily 2008.10615}}].

\bibitem{Bartocci:2024bbf}
R.~Bartocci, P.~B{\"o}er and T.~Hurth, \emph{{Renormalisation group evolution of the shape function g$_{17}$ in $ \overline{B}\to {X}_s\gamma $ and $ \overline{B}\to {X}_s{\ell}^{+}{\ell}^{-} $ at subleading power}}, \href{https://doi.org/10.1007/JHEP04(2025)066}{\emph{JHEP} {\bfseries 04} (2025) 066}, [\href{https://arxiv.org/abs/2411.16634}{{\ttfamily 2411.16634}}].

\bibitem{Schwinger:1960qe}
J.~S. Schwinger, \emph{{Brownian motion of a quantum oscillator}}, \href{https://doi.org/10.1063/1.1703727}{\emph{J. Math. Phys.} {\bfseries 2} (1961) 407--432}.

\bibitem{Keldysh:1964ud}
L.~V. Keldysh, \emph{{Diagram Technique for Nonequilibrium Processes}}, \href{https://doi.org/10.1142/9789811279461_0007}{\emph{Sov. Phys. JETP} {\bfseries 20} (1965) 1018--1026}.

\bibitem{Beneke:2022msp}
M.~Beneke, P.~B\"oer, J.-N. Toelstede and K.~K. Vos, \emph{{Light-cone distribution amplitudes of heavy mesons with QED effects}}, \href{https://doi.org/10.1007/JHEP08(2022)020}{\emph{JHEP} {\bfseries 08} (2022) 020}, [\href{https://arxiv.org/abs/2204.09091}{{\ttfamily 2204.09091}}].

\bibitem{Beneke:2019slt}
M.~Beneke, C.~Bobeth and R.~Szafron, \emph{{Power-enhanced leading-logarithmic QED corrections to $B_q \to \mu^+\mu^-$}}, \href{https://doi.org/10.1007/JHEP10(2019)232}{\emph{JHEP} {\bfseries 10} (2019) 232}, [\href{https://arxiv.org/abs/1908.07011}{{\ttfamily 1908.07011}}].

\bibitem{Qin:2022rlk}
Q.~Qin, Y.-L. Shen, C.~Wang and Y.-M. Wang, \emph{{Deciphering the Long-Distance Penguin Contribution to $\bar{B}_{d,s} \to \gamma \gamma$ Decays}}, \href{https://doi.org/10.1103/PhysRevLett.131.091902}{\emph{Phys. Rev. Lett.} {\bfseries 131} (2023) 091902}, [\href{https://arxiv.org/abs/2207.02691}{{\ttfamily 2207.02691}}].

\bibitem{Piscopo:2023opf}
M.~L. Piscopo and A.~V. Rusov, \emph{{Non-factorisable effects in the decays $ {\overline{B}}_s^0\to {D}_s^{+}{\pi}^{-} $ and $ {\overline{B}}^0\to {D}^{+}{K}^{-} $ from LCSR}}, \href{https://doi.org/10.1007/JHEP10(2023)180}{\emph{JHEP} {\bfseries 10} (2023) 180}, [\href{https://arxiv.org/abs/2307.07594}{{\ttfamily 2307.07594}}].

\bibitem{Huang:2023jdu}
Y.-K. Huang, Y.~Ji, Y.-L. Shen, C.~Wang, Y.-M. Wang and X.-C. Zhao, \emph{{Renormalization-Group Evolution for the Three-Particle B-Meson Soft Function}}, \href{https://doi.org/10.1103/PhysRevLett.133.171901}{\emph{Phys. Rev. Lett.} {\bfseries 133} (2024) 171901}, [\href{https://arxiv.org/abs/2312.15439}{{\ttfamily 2312.15439}}].

\bibitem{Boer:2023vsg}
P.~B{\"o}er and T.~Feldmann, \emph{{Structure-dependent QED effects in exclusive $B$-meson decays}}, \href{https://doi.org/10.1140/epjs/s11734-024-01091-9}{\emph{Eur. Phys. J. ST} {\bfseries 233} (2024) 299--323}, [\href{https://arxiv.org/abs/2312.12885}{{\ttfamily 2312.12885}}].

\bibitem{Feldmann:2023plv}
T.~Feldmann and N.~Gubernari, \emph{{Non-factorisable contributions of strong-penguin operators in $\Lambda_b \to \Lambda \ell^+\ell^-$ decays}}, \href{https://doi.org/10.1007/JHEP03(2024)152}{\emph{JHEP} {\bfseries 03} (2024) 152}, [\href{https://arxiv.org/abs/2312.14146}{{\ttfamily 2312.14146}}].

\bibitem{Bosch:2004th}
S.~W. Bosch, B.~O. Lange, M.~Neubert and G.~Paz, \emph{{Factorization and shape function effects in inclusive B meson decays}}, \href{https://doi.org/10.1016/j.nuclphysb.2004.07.041}{\emph{Nucl. Phys. B} {\bfseries 699} (2004) 335--386}, [\href{https://arxiv.org/abs/hep-ph/0402094}{{\ttfamily hep-ph/0402094}}].

\bibitem{toappear}
R.~Bartocci, P.~Böer and T.~Hurth, \emph{{to appear}}, .

\end{thebibliography}\endgroup

\end{document}